\documentclass[pre,reprint,superscriptaddress,reprint,floatfix,longbibliography]{revtex4-1}
\usepackage{graphicx}
\usepackage{times}
\usepackage{physics}   
\usepackage{bm}        
\usepackage{amssymb}   
\usepackage{amsmath}
\usepackage{wasysym,scalerel}
\usepackage{subcaption}
\usepackage{xcolor}
\usepackage{float}
\usepackage{graphicx}
\usepackage{comment} 
\usepackage{caption,tabularx}
\usepackage{booktabs} 

\begin{document}

\title{Tunnelling-induced restoration of  classical degeneracy in quantum kagome ice}
\author{Kai-Hsin Wu}
\affiliation{Department of Physics and Center for Theoretical Physics, National Taiwan University, Taipei 10607, Taiwan}
 
\author{Yi-Ping Huang}
\affiliation{Max Planck Institute for the Physics of Complex Systems, 
D-01187 Dresden, Germany}

\author{Ying-Jer Kao } 
\email{yjkao@phys.ntu.edu.tw}
\affiliation{Department of Physics and Center for Theoretical Physics, National Taiwan University, Taipei 10607, Taiwan}
\affiliation{National Center for Theoretical Sciences, National Tsing Hua University, Hsin-Chu 30013,  Taiwan}
\affiliation{Department of Physics, Boston University, 590 Commonwealth Avenue, Boston, Massachusetts 02215, USA}
\date{\today}

\begin{abstract}
Quantum effect is expected to dictate the behaviour of physical systems at low temperature. 
For quantum magnets  with geometrical frustration, quantum fluctuation usually lifts the  macroscopic classical degeneracy, and exotic quantum states emerge. 
However, how different types of quantum processes entangle
wave functions in a constrained Hilbert space is not well understood.
Here, we  study  the topological entanglement entropy (TEE) and the thermal entropy of a quantum ice model on a geometrically frustrated kagome lattice.
We find that the system does not show a $Z_2$ topological order down to extremely low temperature, yet  continues to behave like a classical kagome ice with finite residual entropy.
 Our theoretical analysis indicates an intricate competition of  off-diagonal and diagonal quantum processes leading to the quasi-degeneracy of states and effectively, the classical degeneracy is restored.
\end{abstract}

\maketitle

In systems with macroscopic ground state degeneracy, quantum correlation  introduces non-trivial constraints on the Hilbert space, leading to the emergence of highly entangled quantum states of matter.
The scenario is the gist in the studies of quantum Hall effect~\cite{QHE}, flat band
physics~\cite{Tang2011,Neupert2011,Sun2011} and quantum spin 
liquids~\cite{Anderson1973,Anderson1987,Moessner2001,gingras2014,2017ZhouRev,savary2016,knolle2018}. 
Among them, quantum magnets with geometrical frustration have become a fruitful playground to search for exotic quantum phases. 
In particular, spin ice systems on  the corner-sharing tetrahedron lattices have
attracted  enormous attention due to their relevance to rare-earth pyrochlore
materials~\cite{gingras2014,2010GardnerRev,savary2016,knolle2018} and the possibility
to explore exotic quantum states of matter with anisotropic quantum
exchange~\cite{Hermele2004,Savary2012,Lee2012,Shannon:2012ql,Huang:2014uq,Kato:2015uh,Li:2016ai,Li:2017qr,Savary2017}.

Strong spin-orbit couplings in these materials lead to relatively unexplored anisotropic quantum effects.
Dominant ferromagnetic Ising coupling in pyrochlore spin ice materials aligns spins along the
local $\langle 111\rangle$ directions on the tetrahedron, and the system becomes geometrically frustrated at low temperatures.
The macroscopically degenerate ground states obey the so-called ``ice rules'',
with two spins pointing in and two spins pointing out of the centre of each tetrahedron.
By introducing different quantum tunnelling processes, it is possible to drive spin ice systems into various exotic quantum phases~\cite{Hermele2004,Savary2012,Shannon:2012ql,Lee2012,Savary2017}.

\begin{figure*}[ptb]
	
	\includegraphics[width=0.6\linewidth]{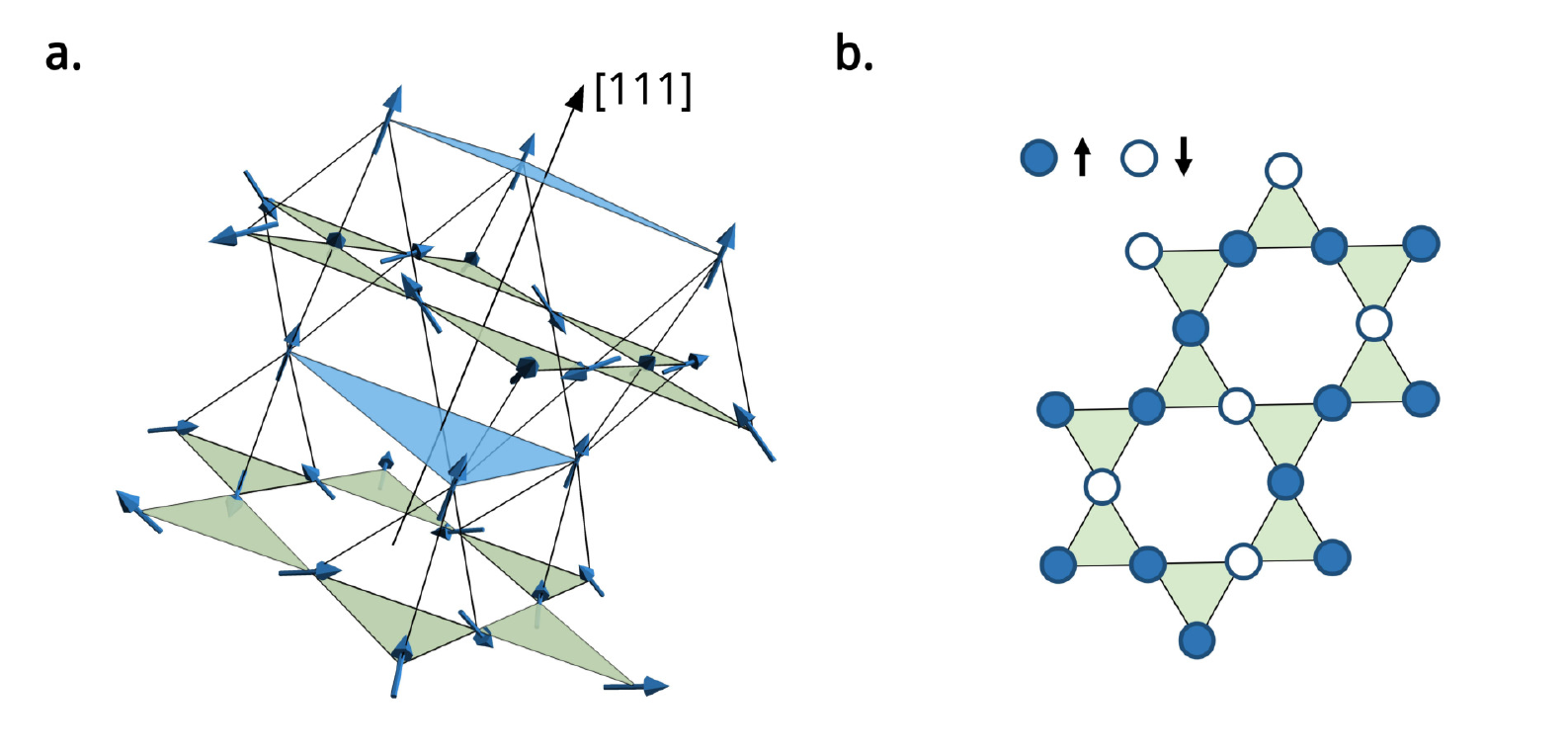}
	\caption{ \textbf{Geometry of pyrochlore lattice and the kagome layer}. \textbf{a.} A pyrochlore spin ice consists of corner sharing tetrahedra of spins pointing into or out  of each tetrahedron. The ferromagnetic coupling between nearest-neighbour spins leads to geometrical frustration, where spins on each tetrahedron follow a '2-in-2-out' ice rule. Pinning the spin on the triangular layers by applying a strong enough field along the [111] direction reduces the pyrochlore lattice to decoupled kagome layers (green layers). \textbf{b.} Kagome lattice can be viewed as corner-sharing triangles, or equivalently, corner-sharing hexagons. 
		In a field, pseudo-spins on each triangle satisfies the kagome ice rule with '2-up-1-down' or '2-down-1-up' depending on the sign of the field ( $h > 0$ here). 
	}
	\label{Fig:kagome} 
\end{figure*}

In addition to the intriguing physics in three dimensions, these pyrochlore spin ice materials also serve as a playground for studying quantum ice physics on a kagome lattice.
The pyrochlore lattice can be visualized as alternating layers of triangular and kagome lattice stacking along the [111] direction (Fig.~\ref{Fig:kagome}a).
When an external field along this axis pins the spins on the triangular layer,
effectively the system becomes decoupled layers of two-dimensional (2D) kagome lattices, provided the field is not too strong.
This dimensional reduction partially reduces the degeneracy, and the ice rule is modified to the kagome ice rule with two spins pointing
into each triangle,  and  one out (2-up-1-down in terms of the pseudo-spin), or vice
versa, as shown in Fig.~\ref{Fig:kagome}b.

Recently, numerical simulation on the kagome lattice that focuses on the pair-flipping process finds a gapped disordered quantum state, dubbed as quantum kagome ice
(QKI) ~\cite{Hao:2015qki}, which is argued to be an exotic $Z_2$ QSL. However, direct
evidence characterising the non-trivial entanglement pattern in the ground state, such
as the TEE, has not been analysed. 
Furthermore, how pair flipping processes induce quantum effects to the ice manifold is not clear.
Using large-scale quantum Monte Carlo (QMC) simulations and degenerate perturbation theory (DPT), we show that the QKI state does not show a $Z_2$ topological order, but continues to behave like a classical kagome ice (CKI) due to the competition among different quantum tunnelling processes. 
Such competition originating from anisotropic exchange coupling could be relevant for pyrochlore
material and recently synthesized tripod Kagome material~\cite{scheie2018crystal}.

\section{Results}

\textbf{Quantum kagome ice model with pair-flipping interaction}--- On a kagome lattice, the nearest-neighbour, symmetry-allowed exchange interactions for the ground state 
dipolar-octupolar doublets can be modelled with an effective pseudo-spin-1/2 XYZh model~\cite{Hao:2015qki},
\begin{widetext}
	\begin{equation}
	H_{\textrm{XYZh}} = \sum_{\langle \mathbf{r},\mathbf{r}' \rangle} J_z S^{z}_{\mathbf{r}} S^{z}_{\mathbf{r}'} - h\sum_{\mathbf{r}} S^z_{\mathbf{r}}- \sum_{\langle \mathbf{r},\mathbf{r}' \rangle} \frac{J_{\pm}}{2} \left( S^{+}_{\mathbf{r}} S^{-}_{\mathbf{r}'} + S^{-}_{\mathbf{r}}  S^{+}_{\mathbf{r}'} \right) + \frac{J_{\pm\pm}}{2} \left( S^{+}_{\mathbf{r}} S^{+}_{\mathbf{r}'} + S^{-}_{\mathbf{r}}  S^{-}_{\mathbf{r}'} \right) 
	\label{XYZh}
	\end{equation}
\end{widetext}
where  $J_z>0$, $\mathbf{r}$  labels kagome lattice sites, and $\langle
\mathbf{r},\mathbf{r}' \rangle$ denotes the nearest-neighbour pairs. 
The first two terms correspond to the CKI model in a field, the third term corresponds to the hopping exchange and the last term is the pair-flipping interaction. 
We emphasize that even though the model is derived from the dipolar-octupolar doublets, the anisotropic exchange is ubiquitous in related materials, and we focus on the simplest  anisotropic exchange term, $S^+_{\boldsymbol{r}}S^+_{\boldsymbol{r}'}$, in the system that can be simulated with large-scale QMC.
In the following, we set $J_z = 1$ unless explicitly stated otherwise. 

For $J_{\pm\pm} = 0$ and $J_{\pm} > 0$, this model is equivalent to the XXZ model with an external field. 
Previous studies show the ground state as a valence-bond solid (VBS) phase with a three-fold degeneracy~\cite{Isakov:2006vbs,Kedar:2006vbs}.
With $J_{\pm} = 0$ and $J_{\pm\pm} < 0$, recently the model is proposed to harbour a $Z_2$ QSL both numerically~\cite{Hao:2015qki} and theoretically~\cite{Huang:2017fk}.
Note that the parameter space of $J_{\pm\pm} > 0$ and $J_{\pm\pm} < 0$ are physically equivalent, connected via a unitary transformation  $S^+\to iS^+$.
Without loss of generality, here, we analyse  model (\ref{XYZh}) with $J_{\pm\pm}<0$. 

\textbf{Topological entanglement entropy}---
For a quantum system with short-range interaction, the Renyi entanglement entropy between subregion $A$ and its complement obeys the so-called \textit{area law},		
\begin{equation}
	S_n(A)= \kappa l - \eta \gamma   + O(L^{-1}),
\end{equation} 
where $\kappa$ is a non-universal constant and $l$ is the boundary length of the subregion. 
$\gamma$ is  the TEE and  $\eta$ is related to the number of (disconnected) boundaries.
 As a universal constant, the TEE plays the role of   ``order parameter'' for detecting the hidden topological order in the system~\cite{Kitaev:2006tee,Wen:2006topo}.
The value of the TEE is related to the quantum dimension $\mathcal{D}$ with $\gamma = \ln \mathcal{D}$ that
characterises the quasi-particle fractionalization of the topological
order~\cite{Kitaev:2006tee}. 
For a system with $Z_2$ topological order, the quantum dimension $\mathcal{D}=2$, and 
$\gamma = \ln 2$ is expected ~\cite{Rowell2009,Jiang:2012zr}.

\begin{figure*}[ptb]
\begin{minipage}{0.8\linewidth}
\includegraphics[width=\linewidth]{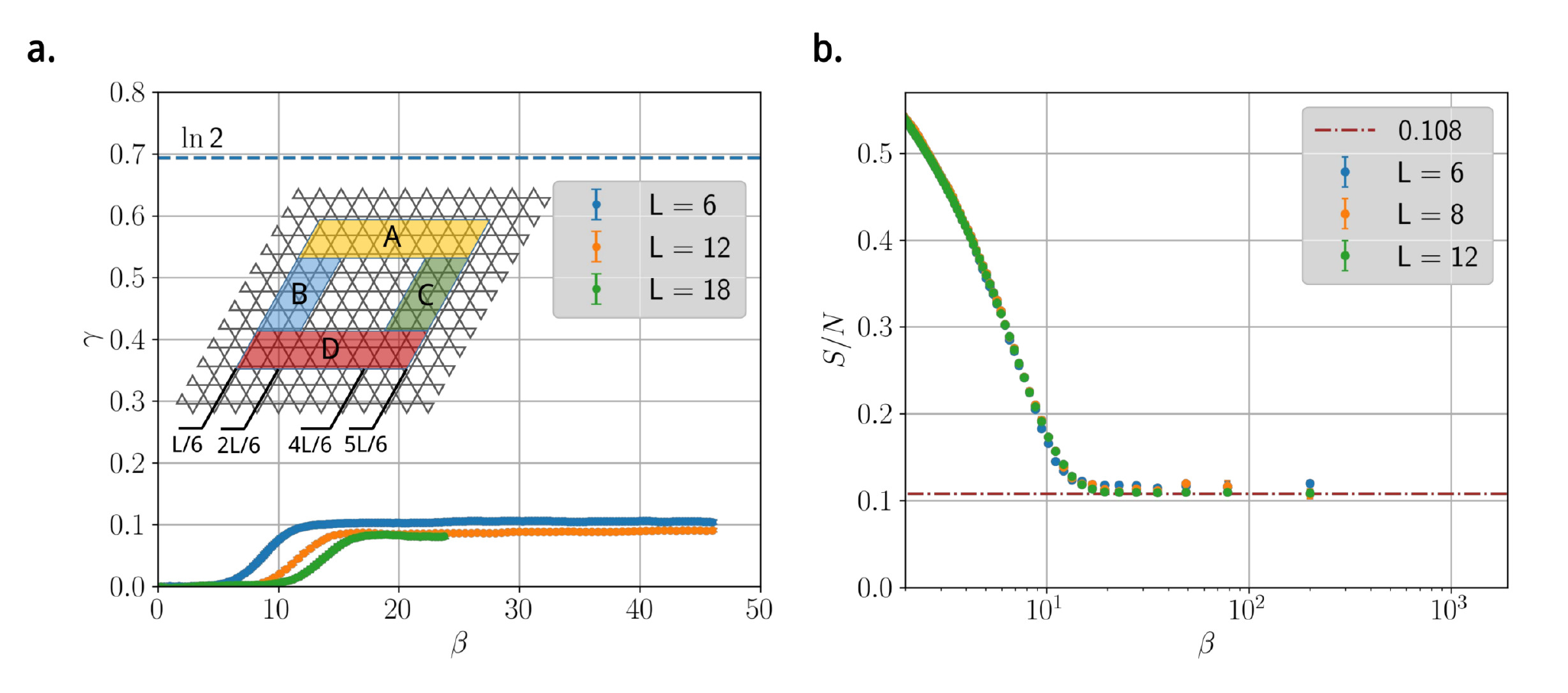}
\end{minipage}
\caption{ \textbf{Topological entanglement entropy and thermal entropy.} 
\textbf{a.} Topological entanglement entropy (TEE) as a function of the inverse temperature $\beta=1/T$  with parameters in the QKI phase ($J_{\pm\pm} = -0.49$ , $J_{\pm} = 0$, and $h = 0.833$). 
The system size is $N$=3$ \times L \times L$.
$\gamma$ converges to a value far smaller than $\ln 2 $ (blue horizontal line), indicating the system does not show a $Z_2$ topological order. 
Four regions in Eq.~(\ref{eq:TEE}) are defined as $R_1 \equiv A \cup B \cup D$, $R_2 \equiv A \cup C \cup D$, $R_3 \equiv A \cup D$ and $R_4 \equiv A \cup B \cup C \cup D$.
\textbf{b.} Thermal entropy per spin $S/N$ as a function of $\beta$  in the QKI phase ( $J_{\pm\pm} = -0.45$ , $J_{\pm} = 0$ , $h = 1$). 
The low temperature plateau is consistent with the residual entropy per spin of a CKI (brown horizontal dot-dashed line). 
}
\label{Fig:Data} 
\end{figure*}

 We measure the quantum entanglement using the second  Renyi entropy~\cite{Isakov:2011ee,Poilblanc:2011cy}, 
\begin{align}
			S_2(A) = -\ln  \textrm{Tr}({\rho_A^2}),
\end{align} 
where $\rho_A$ is the reduced density matrix of subregion $A$.
Using the replica trick~\cite{Isakov:2011ee}, we measure  $S_2$ with four different
subregions (Fig.~\ref{Fig:Data}a) that are strategically designed to eliminate the area terms~\cite{Wen:2006topo}, and
\begin{equation}
2\gamma = -S_2(R_1) - S_2(R_2) + S_2(R_3) + S_2(R_4).
\label{eq:TEE}
\end{equation}

Fig.~\ref{Fig:Data}a shows  TEE as a function of the inverse temperature $\beta=1/T$ with  parameters in the QKI regime ($J_{\pm\pm} = -0.49$, $J_{\pm} = 0$, and $h = 0.833$).
We find $\gamma$ is far below the expected  $\ln2$ value even at a temperature as low as $T\approx 1/48 (\beta=48)$, indicating  the system does not have a $Z_2$ topological order. 
The small finite $\gamma$ at low temperature  is due to sub-leading corrections that cannot be cancelled.  
This result suggests two possibilities: either the QKI state is a short-range entangled symmetry
protected topological order or the quantum fluctuations couple different kagome ice states in such
a manner that the system behaves classically. 
We  clarify this issue through the study of thermal entropy at low temperature.

\textbf{Thermal entropy}---  The ground states of the XYZh model (\ref{XYZh})  in the
classical limit  satisfy the 2-up-1-down kagome ice rule and are extensively
degenerate, leading to a residual entropy per spin
$S/N=0.108$~\cite{MoessnerIQF}.

In order to directly measure the thermal entropy in our QMC simulations, we employ the Wang-Landau method~\cite{Troyer:2004WL}.
We observe the thermal entropy remains finite at an extremely low temperature  $T=1/200 (\beta=200)$ with the value corresponding to the residual entropy per spin of a CKI. 
This classical behaviour in the supposedly quantum region is counter-intuitive.
The system neither enters an ordered phase through the quantum order-by-disorder
scheme~\cite{villain_ObD,Savary2012b,Zhitomirsky2012} nor becomes a highly entangled disordered quantum state.
To solve this puzzle, we analyse  possible quantum processes out of the CKI manifold
using  DPT~\cite{Bergman2007a,Bergman2007b}.

\begin{figure*}
	\begin{minipage}{0.75\linewidth}
		\includegraphics[width=\linewidth]{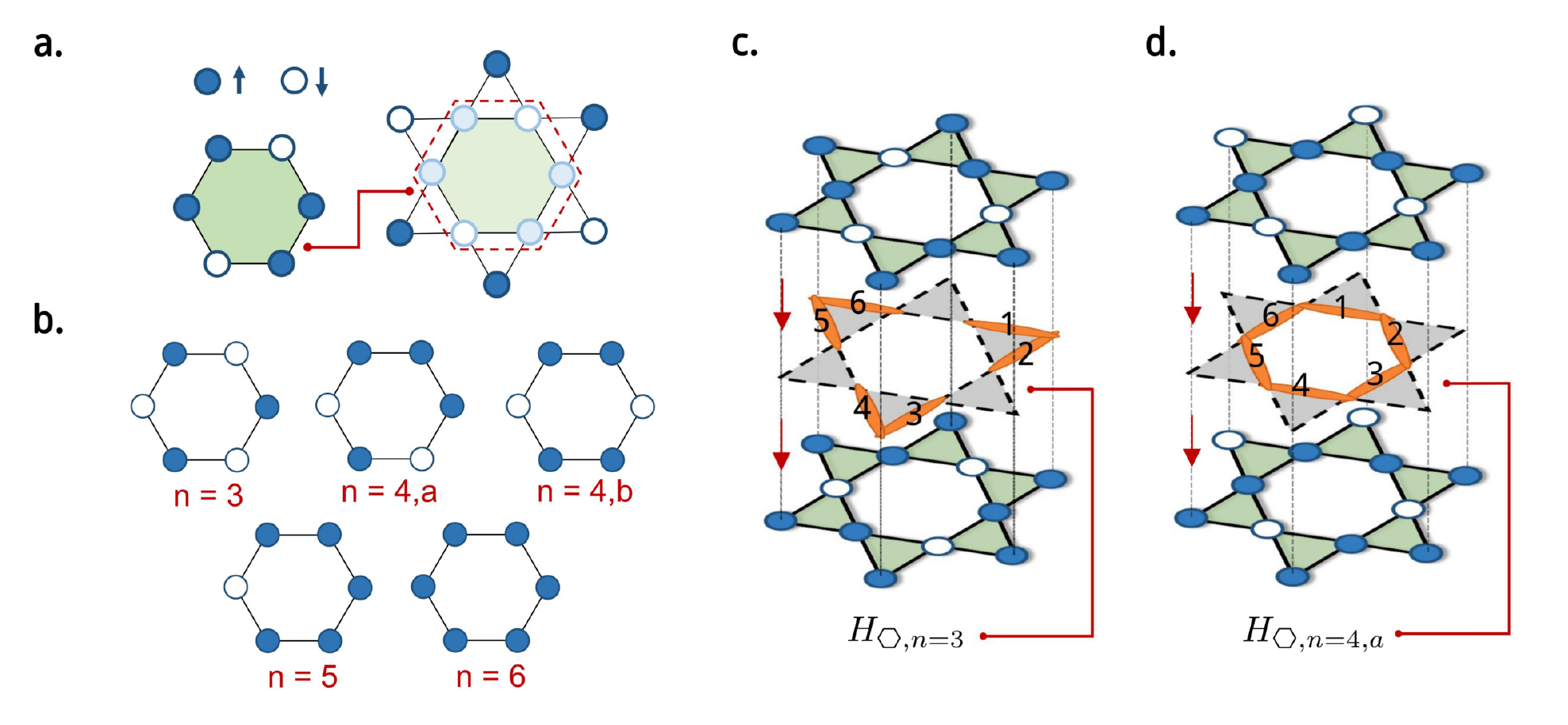}
	\end{minipage}
	\caption{ \textbf{Hexagon units and quantum tunnelling processes.} \textbf{a.} In a  field, spins on each triangle follow the 2-up-1-down rule. 
		When the spin configuration on the hexagon is determined, the configuration of the remaining spins within the star of David is uniquely defined. 
		\textbf{b.} The allowed hexagon configurations in a field, classified according to the number of up-spins $n$. 
		\textbf{c.} Example of the off-diagonal tunnelling term  $H_{\hexagon,n=3}$ formed by six nearest-neighbour $S^{\pm}S^\pm$  operators on the orange bonds that couples two different $n=3$ CKI states  .
		\textbf{(d)}  Example of the diagonal tunnelling term $H_{\hexagon,n=4,a}$  that couples the same $n=4,a$ CKI states. 
	}
	\label{Fig:DPT} 
\end{figure*}

\textbf{Degenerate Perturbation Theory}--- Starting from the classical model, we treat all the quantum fluctuations as perturbations. 
It is useful in the following analysis to view the kagome lattice   as corner-sharing hexagons; thus, all the  non-trivial perturbation processes are  within a single star of David (Fig.~\ref{Fig:DPT}a). 
Due to the presence of the field that splits the degeneracy of the kagome ice rule on each triangle, the 2-up-1-down configurations are favoured. 
Spin configuration on each star is therefore uniquely determined by the hexagon configuration (Fig.~\ref{Fig:DPT}b) that determines  the process in the perturbation theory. 

First, we consider the case $J_{\pm\pm} \neq 0$ and $J_{\pm} = 0$ where the
proposed QKI  is realized. 
The leading non-trivial processes appear at the sixth-order
of  perturbation, and an effective Hamiltonian $\hat{P}_6$ can be written as,
\begin{align}
\hat{P}_6 &= H_d + K_{pp} \sum_{\forall \hexagon n=3 } H_{\hexagon,n=3} \\
H_d &=   D_{4,a}\sum_{\forall \hexagon n=4,a } H_{\hexagon,n=4,a}  
+D_{4,b}\sum_{\forall \hexagon n=4,b }  H_{\hexagon,n=4,b} \nonumber\\ 
&+ D_5 \sum_{\forall \hexagon n=5 } H_{\hexagon,n=5} \nonumber\\
&+ D_6 \sum_{\forall \hexagon n=6 } H_{\hexagon,n=6}.
\label{eq:P6}
\end{align}
$H_{\hexagon,n=3}$  acts on an $n$=3 hexagon to generate an effective ring-exchange process, as shown in Fig.~\ref{Fig:DPT}c, which brings one CKI configuration to a different one.
In addition to the ring-exchange term, various  non-trivial diagonal terms $H_d$ appear at the same order of perturbation acting on hexagons $n\ge 4$. For the case of $n=4$, there exist two different terms $H_{\hexagon,n=4,a}$ and $H_{\hexagon,n=4,b}$ acting on the  $a$ and $b$ types of hexagons (Fig.~\ref{Fig:DPT}b) respectively. 
The coefficients of these processes can be directly computed in  DPT (See Supplementary Information for details). 
In the case of $h=J_z$, we have $K_{pp}=-\frac{58}{81}\Gamma$, $D_{4,b}=-\frac{1}{6}\Gamma$, $D_{4,a}=-\frac{1}{36}\Gamma$, $D_5=-\frac{289}{5292}\Gamma$, $D_6=-\frac{2}{49}\Gamma$, where   $\Gamma={J_{\pm\pm}^6}/{J_z^5}$.

Consider the other limiting case where  $J_{\pm} \neq 0$ and $J_{\pm\pm} = 0$. The lowest non-trivial process occurs at the third order of perturbation, with an effective Hamiltonian, 
\begin{equation}
\hat{P}_3  = K_{np} \sum_{\forall \hexagon n=3 } H_{\hexagon,n=3} + c, 
\end{equation} 
where $K_{np} = -12{J^3_{\pm}}/{J_z^2}$. 
All diagonal processes  at this level contribute to an overall constant energy shift $c$ which is irrelevant. 
The ring-exchange term $H_{\hexagon,n=3}$  drives the system into a VBS ground state~\cite{Isakov:2006vbs,Kedar:2006vbs}.

\begin{figure*}[ptb]
\begin{minipage}{0.8\linewidth}
\includegraphics[width=\linewidth]{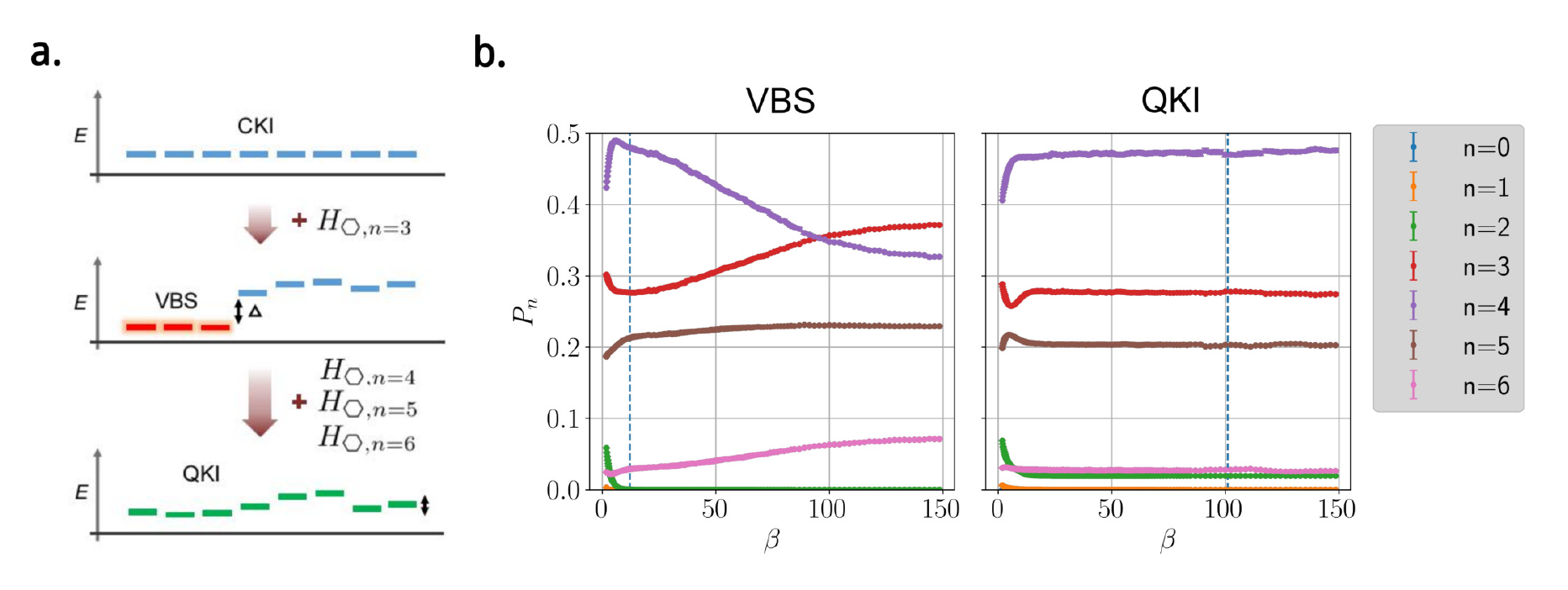}
\end{minipage}
\caption{ \textbf{Effects of  quantum tunnelling processes}. \textbf{a.} The schematic picture
of the energy level reorganization due to  the sixth-order perturbation  (\ref{eq:P6}). 
Starting from the degenerate classical ice manifold, we first introduce the ring-exchange term $H_{\hexagon,n=3}$. 
This will select the three-fold degenerate VBS states with an energy gap $\Delta$. 
Further adding  the diagonal terms $H_{\hexagon,n=4,5,6}$,  the energy levels are
reorganized to become  quasi-degenerate with a suppressed energy gap. 
\textbf{b.} QMC results of  the hexagon fraction  $P_n$  v.s. $\beta$  in the VBS regime with $J_{\pm}=0.19$ and $J_{\pm\pm} = 0$ (left panel); and in the QKI regime $J_{\pm\pm} =-0.49$ and $J_{\pm}=0$ (right panel). Both are under a field $h = J_z$.
The vertical dashed lines indicate the perturbative energy scale estimated by the leading ring-exchange contribution with $\beta \sim 12.1/J_z$ and $\beta \sim 100.9/J_z$ for the left and right panel respectively. 
	}    
	\label{Fig:Elevl} 
\end{figure*}

Fig.~\ref{Fig:Elevl}a shows a schematic picture to illustrate the effects coming from the diagonal and off-diagonal quantum tunnelling processes.  
For a QKI  ($J_{\pm\pm} \neq 0$ and $J_{\pm}=0$), the introduction of the off-diagonal ring-exchange  $H_{\hexagon,n=3}$  selects the three-fold degenerate VBS state out of the degenerate classical ice manifold, leaving all  other states at higher energies. 
Adding the diagonal terms, reconfiguration of the energy levels occurs.
These terms  tend to maximize the overall fraction of $n$=4, 5, 6
hexagons while minimizing the fraction of $n$=3 hexagons.
The competition between diagonal and off-diagonal processes reorganizes the states into quasi-degenerate levels and the classical degeneracy is restored. 
On the other hand, for the case $J_{\pm}\neq 0$ and $J_{\pm\pm} = 0$, the process terminates at the ring-exchange level, and the VBS ground state is selected.

We further demonstrate this mechanism by  measuring $P_n$, the fraction of hexagons with $n$ up spins using QMC (Fig.~\ref{Fig:Elevl}b).
For the VBS parameters,  the weight of $n$=3 hexagons  $P_3$ dramatically increases at low temperature,  accompanied with the decrease of $P_4$ . 
On the other hand, for the QKI parameters, the fractions for each type of hexagons remain  unchanged, suggesting the system remains within the CKI state down to temperature much lower than the perturbative energy scale. 
Although DPT is  expected to work only in the small $J_{\pm\pm}/J_z$ limit,  the QMC results indicate the competition between these quantum tunnelling processes is indeed nonperturbative.

To illustrate the quantum origin of this quasi-degeneracy, we  study the effective model using exact diagonalisation.
Here, we slightly modify the effective Hamiltonian by introducing a tuning parameter $\alpha$ in order to change the weight of the diagonal process,     
\begin{equation}
	\hat{P}_6(\alpha) = \alpha H_d + K_{pp} \sum_{\forall \hexagon n=3 } H_{\hexagon,n=3}. 
	\label{P6p}
\end{equation} 
Since the exact weight ratio between the two types of processes in the original XYZh model is unknown, tuning $\alpha$ provides information for how the energy spectrum is affected by adding the diagonal term. 

For  $\alpha=0$ where the ring-exchange dominates, the ground state should be the three-fold degenerate VBS state. Due to the finite size effect, the three lowest energy states in our ED results are not exactly degenerate.
However, a detailed analysis of the wave function confirms  these states correspond to the VBS state and becomes degenerate in the thermodynamic limit (See Supplementary Information).
For $\alpha\to\infty$, the model corresponds to keeping only the diagonal terms. 
Therefore, all the classical kagome ice configurations are eigenstates of the Hamiltonian.
We find the ground states are also three-fold degenerate, and corresponds to the three charge-ordered states  in the classical kagome ice~\cite{Chern:2011fv,Wolf:CIKag}.

\begin{figure*}[ptb]
\begin{minipage}{0.9\linewidth}
\includegraphics[width=\linewidth]{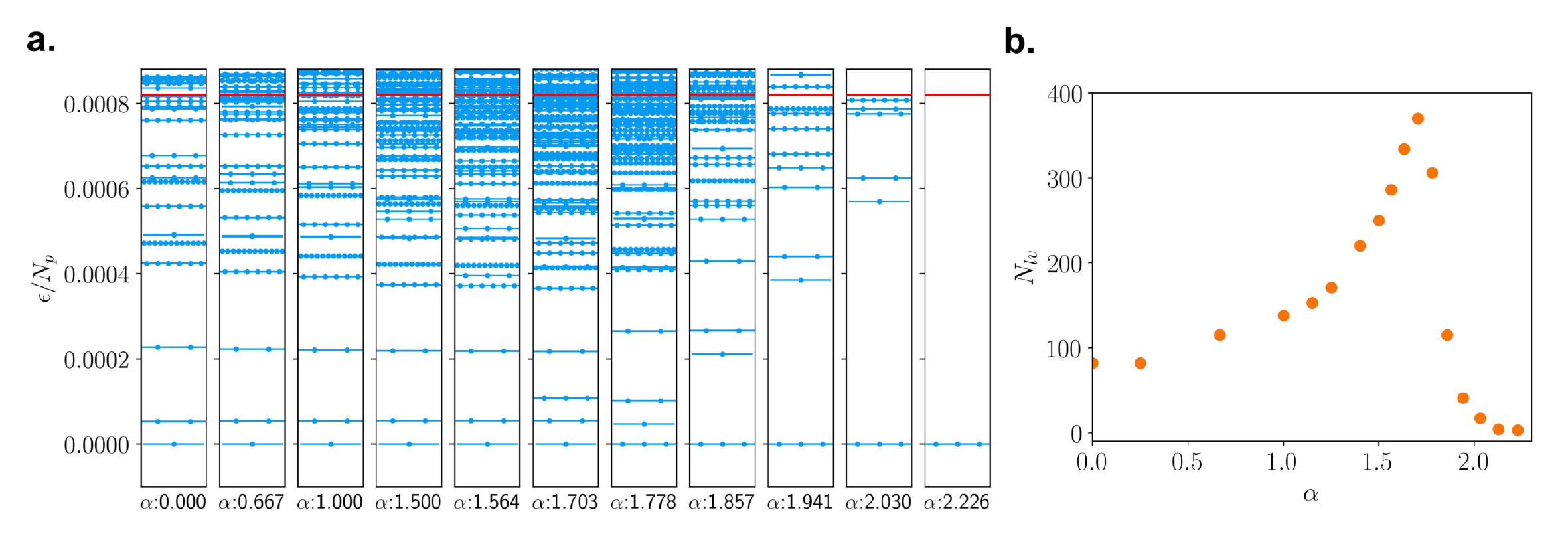}
\end{minipage}
\caption{ \textbf{Spectrum of the effective Hamiltonian and effects of diagonal processes.} \textbf{(a)} The energy spectrum of the effective model $\hat{P_6}$ with tuning parameter $\alpha$. Each dot represents a single state.  The $y$-axis is the energy per-hexagon ($N_p=N/3$ is the number of hexagons), and each ground state is shifted to zero for easy comparison. The red line indicates an energy cutoff $\epsilon/N_p = 0.00082$. \textbf{(b)} Number of energy levels, $N_{lv}$, that lies below the cutoff.  $N_{lv}$ increases as $\alpha$ increases, indicating that the energy spectrum is compressed toward the ground state,   until $\alpha>1.703$  where $N_{lv}$ decreases again.}
	\label{Fig:Spectrum} 
\end{figure*}

We expect there should be a level crossing at some intermediate $\alpha$, which indeed occurs somewhere around  $\alpha=1.703 \sim  1.778$ (Fig.~\ref{Fig:Spectrum}a).
Also, we find that the spectrum is compressed toward the ground state. 
To give a quantitative measure of this compression,  we set an energy cutoff $\epsilon/N_p = 0.00082$ and study how the number of levels below this cutoff, $N_{lv}$,  changes with $\alpha$.
We observe that the $N_{lv}$ increases as $\alpha$ increases, indicating a compression of energy levels toward the ground states, until after $\alpha>1.703$ (Fig.~\ref{Fig:Spectrum}b).
This suggests the quasi-degeneracy observed in our QMC simulation is a consequence of the compressed spectrum due to the competition between diagonal and off-diagonal term.  
With this physical picture in mind, we expect by tuning $K_{pp}$ in the effective Hamiltonian (\ref{eq:P6}),  the VBS phase should emerge with large enough $K_{pp}$. 
This can be realized in the original XYZh model by including both nonzero $J_{\pm}$ and $J_{\pm\pm}$ terms. The emergence of VBS by adding a small $J_{\pm}$ in QKI is then confirmed from the peaks of the static structure factor at VBS ordering momentum vector in our QMC simulation (See Supplementary Information).  

\textbf{Conclusions} 
Although the XYZh model on a kagome lattice has been proposed to be a new playground to search for 2D $Z_2$ QSL,   
our results suggest that the QKI does not show a $Z_2$ topological order down to low temperature, and the system bahaves classically.
The suppression of the quantum energy scale originated from the competition between the off-diagonal ring-exchange and  diagonal processes indicates that a much lower temperature than $T=1/200$
has to be reached before entering the true quantum regime. 
Even if the true quantum ground state is a $Z_2$ QSL with an extremely small gap, it will be very hard to be realized experimentally or confirmed numerically since it is extremely fragile.

Our results also indicate that the kagome ice states cannot be hybridized
easily with quantum anisotropic exchange. Thus, the kagome ice physics is more likely to
be observed at finite temperature experiments with non-trivial dynamics.
This non-perturbative result of QMC provides crucial information for 
understanding the experiments, such as the recent experiments on Nd$_2$Zr$_2$O$_7$~\cite{Lhotel:2018hc}. In addition to pyrochlore materials, such physics could also play a role in the recently synthesized tripod materials~\cite{Paddison:2016fv,scheie2018crystal,dun2018quantum}.

On the other hand, although the true ground state remains unknown, it would be interesting to study the effects of tilting the field away from the [111] axis as this can provide an easy method to tune the weights of the ring-exchange and diagonal processes.
We finish by pointing out the non-trivial diagonal terms we found
through DPT also exist on the pyrochlore lattice since the CKI states are a subset of the ice
manifold.
Further systematic studies are necessary to see if the phenomena discussed in this paper
can be extended to three-dimensional cases \cite{Rau2015,2018:Chen}. 

\section{Methods}
 We  implement the stochastic series expansion (SSE)~\cite{Sandvik91PRB,Syljuasen02PRE}  in the $S^z$ basis with a triangular plaquette break-up of the XYZh Hamiltonian. 
The  directed loop equations are solved using numerical linear solver to minimize the bounce probability. 
The Renyi entanglement entropy is measured by implementing the replica trick~\cite{Isakov:2011ee}. 
In our simulations, we follow the scheme proposed in Ref.~\cite{Melko:2010Renyi} to measure the second  Renyi entanglement entropy $S_2$ with four subregions independently. 
The simulation runs on average $10^8 \sim 10^9$ Monte Carlo steps (MCS) for each  subregion.
The topological entanglement entropy $\gamma$ is  calculated by combining  $S_2$ of  the four subregions with the standard bootstrap resampling procedure.
The thermal entropy is measured  with $10^8$ MCS using SSE with the Wang-Landau algorithm~\cite{Troyer:2004WL} for a long operator string with a fixed length. 
For the exact-diagonalisation of the effective Hamiltonian, we first search for  all basis states that satisfy the 2-up-1-down ice-rule.
We then construct the effective Hamiltonian $\hat{P}_6(\alpha)$, and perform a Lanczos diagonalisation to obtain the energy spectrum and eigenstates. 
The data presented in this paper  requires the computation resources  approximately about  330 CPU core-years on two different heterogeneous high-performance computers (HPCs) with 2.50GHz Intel Xeon or equivalent CPUs at  the National Center for High-performance Computing. 

\begin{acknowledgments}
This work was supported  by the Ministry of Science and Technology (MOST) of Taiwan under Grants No. 105-2112-M-002-023-MY3, and 104-2112-M-002-022-MY3, and was funded in part by a QuantEmX grant 
from ICAM and by the Gordon and Betty Moore Foundation through Grant 
GBMF5305 to Y.J.K. We are grateful to the National Center for High-performance Computing
for computer time and facilities. Y.J.K. thanks Juan Carrasquilla, Mike Hermele and Zi-Yang Meng for useful discussions.
\end{acknowledgments}

\appendix

\section{Degenerate perturbation theory}
	We start by identifying the classical part in the XYZh model as unperturbed system, denoting as $H_0$, and
\begin{align}
	H &= H_0 + V \nonumber\\
	H_0 &= J_z \sum_{\langle i,j\rangle} S^i_{z} S^j_{z} - h\sum_{i} S^{i}_z.
\end{align}
Next, we treat the quantum term $V$ as perturbation acting on the degenerate classical ice manifold $\boldsymbol{\Omega}=\{\Omega_0, \Omega_1, \ldots \}$ with the ice rule "2-up-1-down" ($h>0$) or "2-down-1-up" ($h<0$), 
\begin{align}
	H_0 \boldsymbol{\Omega} &= E_0 \boldsymbol{\Omega}.		
\end{align}
	 Define an operator $\mathbb{P}$ that projects the states $\boldsymbol{\Psi}=\{\Psi_0,\Psi_1, \ldots\}$ in the Hilbert space to the degenerate ice manifold, 
	\begin{align}
	\mathbb{P} \boldsymbol{\Psi} = \boldsymbol{\Omega} 
	\end{align}
	where $\mathbb{P}^2=\mathbb{P}$.

	Following the standard Brillouin-Wigner perturbation theory~\cite{FuldePeter1995Ecim}, the perturbation expansion can be written as~\cite{Bergman2007a,Bergman2007b},
	\begin{align}
	(H_0 + \mathbb{P}    V \sum_{t=0}^{\infty} G^t \mathbb{P} )\boldsymbol{\Omega} &= E \boldsymbol{\Omega}\nonumber \\
	G = \frac{(I-\mathbb{P})}{E - H_0} V. 
	\end{align}
	
	We now have a non-linear eigenvalue problem to solve for the  energy shifts ($\epsilon = E - E_0$), 	
	\begin{align}
	\hat{P} \boldsymbol{\Omega} =\epsilon \boldsymbol{\Omega} 
	\end{align}
	with
	\begin{equation}
	\hat{P} \equiv	\left( V \sum_{t=0}^{\infty} G^t \right).
	\end{equation}
	Essentially, perturbations coming from the quantum fluctuations lift the degeneracy of the ice manifold and quantum phase emerges. 
	In the following calculation, we use hexagon units as defined in the main text and take $h > 0$, where all the triangular plaques follows "2-up-1-down" rule. 

	For the case that the $S^{\pm}S^{\pm}$ is the only present quantum fluctuation, the lowest non-constant term is at the sixth order, 
	\begin{align}
	V G^5 &\equiv \hat{P}_6,
	\end{align}
	which is the sum of off-diagonal $H_o$ and  diagonal $H_d$ contributions. 

\begin{figure*}[ht!]
	\begin{minipage}[b]{0.6\textwidth}
		\includegraphics[width=\textwidth]{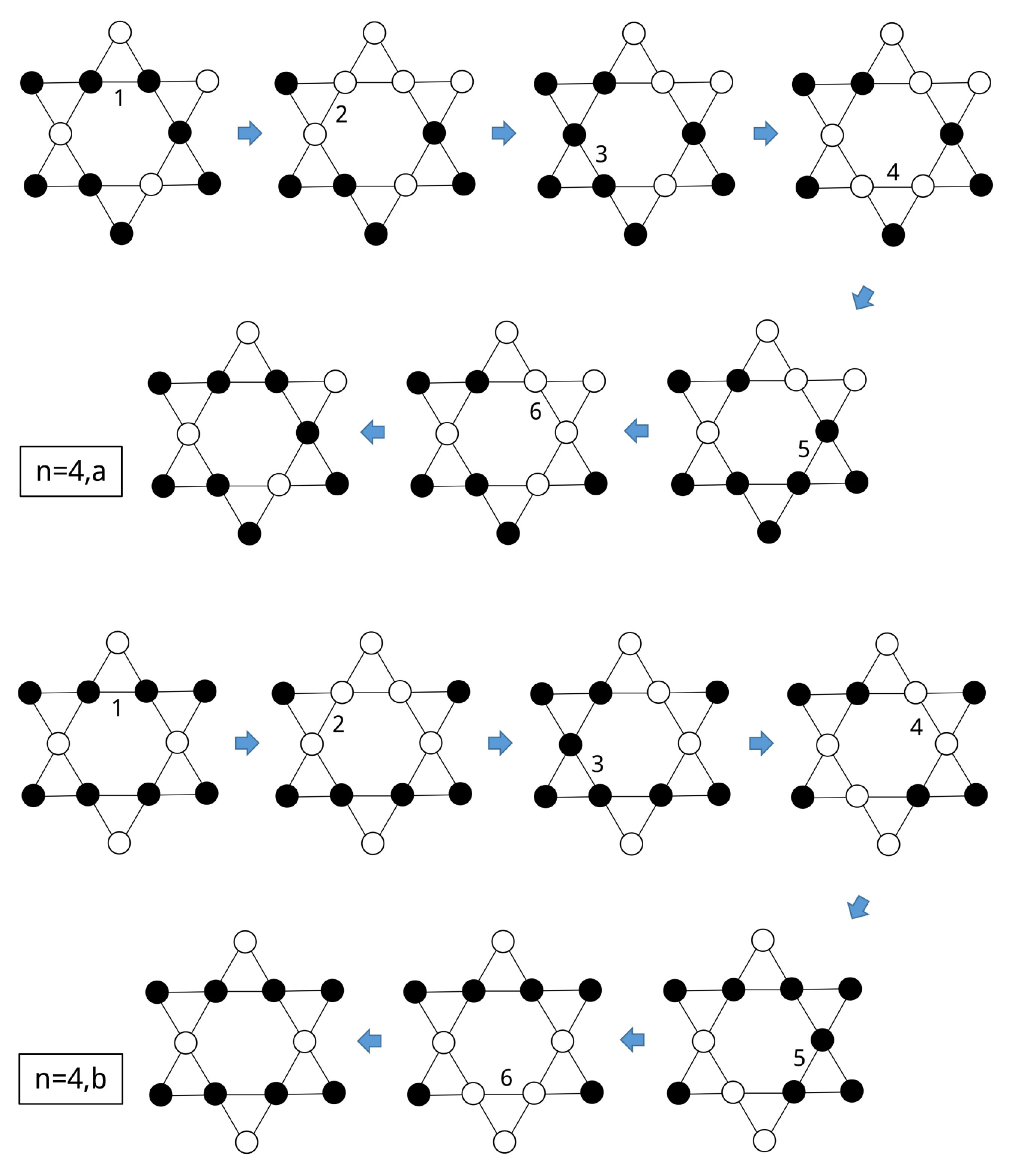}
	\caption{Relevant diagonal processes on two types of hexagon with $n=4$}
	\label{Fig:6thDn4}
	\end{minipage}
\end{figure*}

\begin{figure*}[ht!]	
	\begin{minipage}[b]{0.6\textwidth}
		\includegraphics[width=\textwidth]{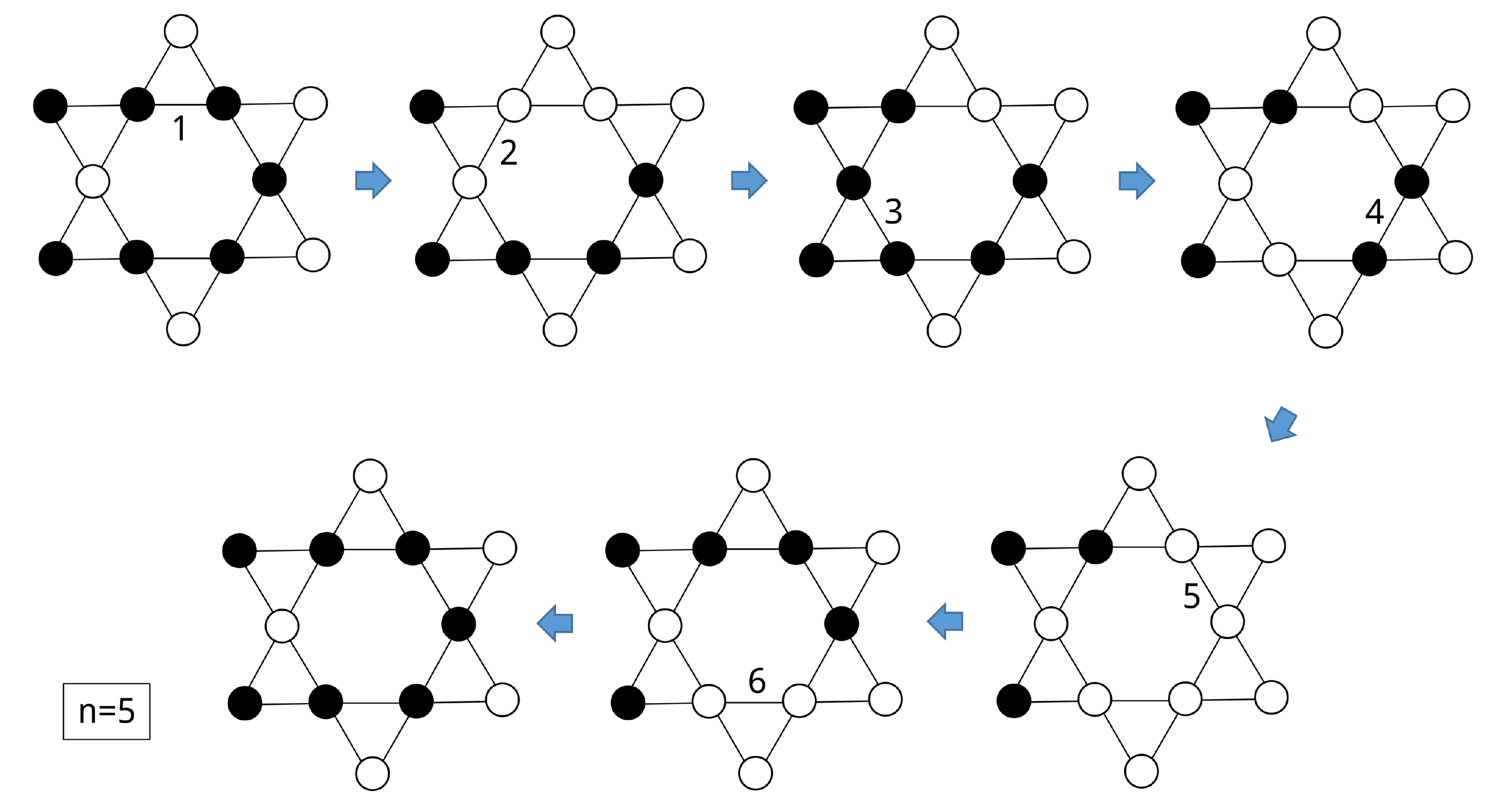}
	\caption{Relevant diagonal process on hexagon with $n=5$}
	\label{Fig:6thDn5}
	\end{minipage}
\end{figure*}

\begin{figure*}[ht!]	
	\begin{minipage}[b]{0.6\textwidth}
		\includegraphics[width=\textwidth]{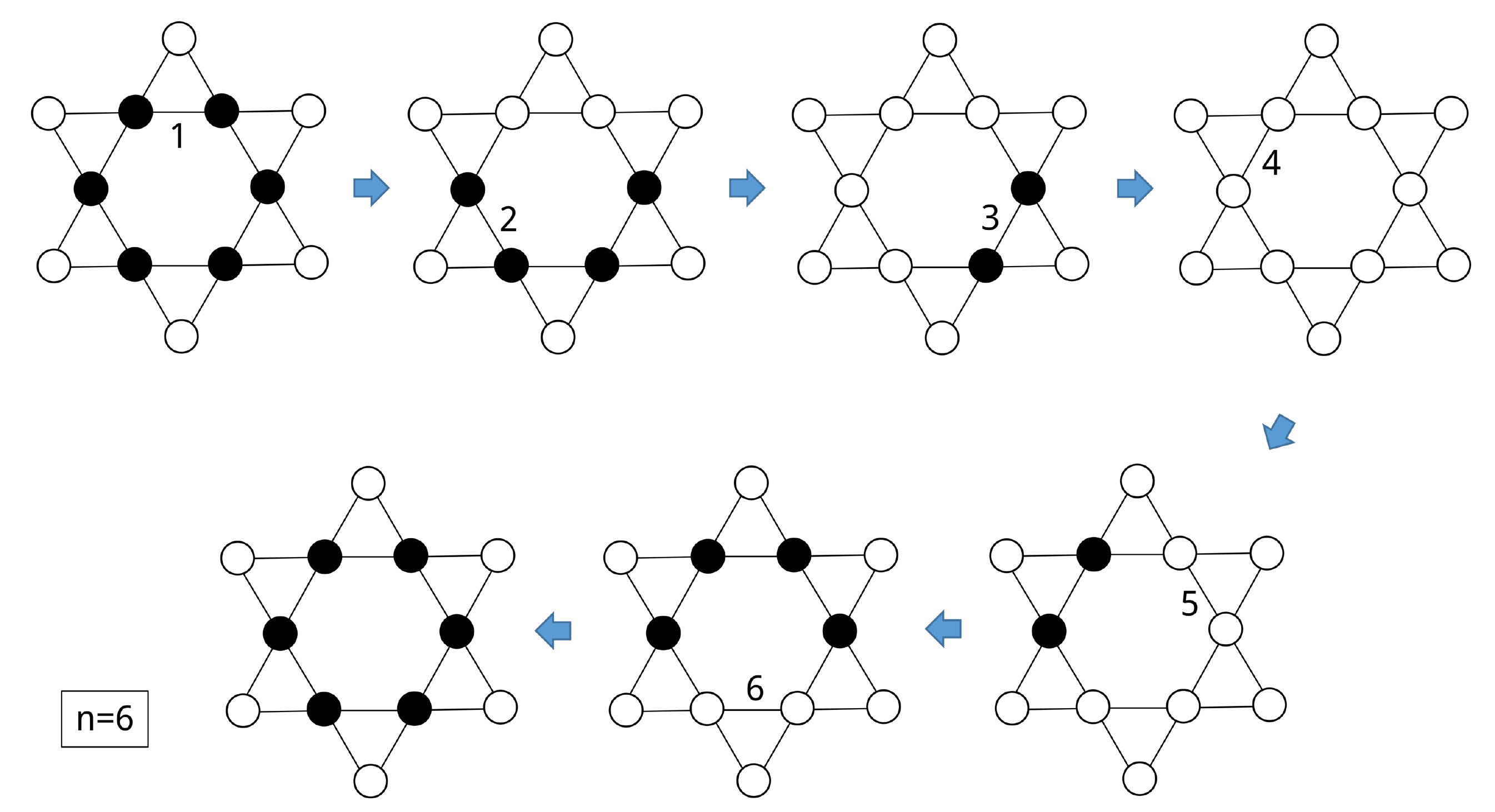}
	\caption{Relevant diagonal process on hexagon with $n=6$}
	\label{Fig:6thDn6}
	\end{minipage}
\end{figure*}

\begin{widetext}
\begin{align}
	\hat{P}_6    &= H_d + H_o \\
	 H_d            &= D_{4,a} \sum_{\forall \hexagon n=4,a } H_{\hexagon} +D_{4,b} \sum_{\forall \hexagon n=4,b } H_{\hexagon} + D_5 \sum_{\forall \hexagon n=5 } H_{\hexagon} + D_6 \sum_{\forall \hexagon n=6 } H_{\hexagon} \label{diag}\\ 
	 H_o            &= K_{pp} \sum_{\forall \hexagon n=3 } H_{\hexagon} \label{offdiag}, 
	\end{align}
\end{widetext}
	where the off-diagonal term corresponds to the ring-exchange process that acts on $n = 3$ hexagons, and 
	the diagonal terms correspond to  processes that  $V$  acts on each bond only once on $n\ge 4$ hexagons   (see Fig.~\ref{Fig:6thDn4}, \ref{Fig:6thDn5}, \ref{Fig:6thDn6}). 
Prefactors associated with each term can be computed by  listing all  possible ways to arrange the local two-site ($S^{\pm}S^{\pm}$ or $S^{\pm}S^{\mp}$) operators ($G$) to form the perturbation operators $H_{\hexagon}$ that transfer  states within the ice manifold~\cite{Bergman2007a,Bergman2007b}.

The prefactors for the off-diagonal ring-exchange term, 
\begin{align}
K_{pp} = -\frac{6J^6_{\pm\pm}}{J_z^2(2h+J_z)^5} \left[ 7J_z^2 + 14J_zh + 8h^2\right]
\end{align}
and  diagonal terms
\begin{align}
	D_{4,a} &= -\frac{J^6_{\pm\pm}}{4hJ_z^2(J_z+2h)^2} \nonumber \\
	D_{4,b} &= -\frac{(2J+h)J^6_{\pm\pm}}{4hJ_z^2(J_z+2h)^2(J-\frac{h}{2})} \nonumber \\
	D_5 &= -J^6_{\pm\pm}\frac{8J(3J_z+4h)(2J_z+h) + h(7J_z+4h)^2}{4hJ_z^2(J_z+2h)^2(2J_z+h)(3J_z+4h)^2}\nonumber\\
	D_6 &= -\frac{12J^6_{\pm\pm}}{(3J_z+4h)^2(J_z+2h)(J_z+h)h}. 
\end{align} 
	

	If we further let $h=J_z$ provided the system is within the lobe, we obtain,
	\begin{align}
	K_{pp}   &= -\frac{58}{81}\frac{J^6_{\pm\pm}}{J_z^5}\nonumber \\
	D_{4,a} &= -\frac{1}{36}\frac{J^6_{\pm\pm}}{J_z^5} \nonumber\\
	D_{4,b} &= -\frac{1}{6}\frac{J^6_{\pm\pm}}{J_z^5} \nonumber\\
	D_5 &= -\frac{289}{5292}\frac{J^6_{\pm\pm}}{J_z^5} \nonumber\\
	D_6 &= -\frac{2}{49}\frac{J^6_{\pm\pm}}{J_z^5}. 
	\end{align}

\section{Thermal entropy measurement using Wang-Landau method }

In general, one can estimate thermal entropy by numerically integrating the specific heat data from QMC.
However, this approach requires a very accurate estimate of the specific heat and suffers from the error due to the discretized  temperature intervals. 
Instead, we use the Wang-Landau sampling scheme~\cite{Troyer:2004WL}  to directly access the thermal entropy in our QMC simulations. 

In the SSE formalism, the partition function is written as
	\begin{align}
	Z &= \textrm{Tr}{ \left[ e^{-\beta H}\right]}\nonumber \\
	&= \sum_{n} \frac{(\beta)^n}{n!} \sum_{\phi,a,b} \bra{\phi} H_{a_0,b_0} ... H_{a_n,b_n}\ket{\phi}\nonumber \\ 
	&=  \sum_{n} \beta^n S_n\nonumber\\
	& =\sum_n W(n),
	\label{eq:Z}
	\end{align}
	where $H_{a_i,b_i}$ is the local Hamiltonian.
In our simulation, we perform triangle plaquette decomposition  of the Hamiltonian as discussed in Ref.~\cite{Melko:2007pd} and sampling using the directed loop algorithm~\cite{Syljuasen02PRE}. 
Here, we rewrite Eq.~(\ref{eq:Z}) into a  generalized representation by introducing a weighting factor $g(n)$, 
\begin{align}
Z' &= \sum_{n} \beta^n S_n\ g(n)\nonumber \\
	&= \sum_{n} W'(n).
\end{align}	
In the simulation, we first search for $g(n)$ such that the modified weight $W'(n)$ are roughly equal, and then sample  $Z'$ with the modified weight $W'(n)$. 

The partition function for a range of arbitrary temperatures $\bar{\beta}$ can be calculated by,
\begin{align}
Z(\bar{\beta}) = \sum_{n} \left(\frac{\bar{\beta}}{\beta}\right)^{n} \frac{W'(n)}{g(n)}. 
\end{align}
	
In our simulation, we fix $\beta=1$ for convenience. 
To obtain the estimates for physical observables,  we first record the estimates for each observables in each $n$ separately, 
\begin{align}
 \langle O_n \rangle &= \sum_{n'} O \delta_{n,n'} \frac{W'(n')}{Z'}\nonumber \\
 \langle I_n \rangle &=  \delta_{n,n'} \frac{W'(n')}{Z'}.
\end{align}
	We then reweight  the estimates with the set $\{g(n)\}$ with an undetermined normalization constant $A$,	
	\begin{align}
	\langle \bar{O}(\bar{\beta}) \rangle &= A\sum_{n}  \frac{(\bar{\beta})^{n}}{g(n)} \langle O_n \rangle \nonumber \\
	\langle \bar{Z}(\bar{\beta}) \rangle &= A\sum_{n}  \frac{(\bar{\beta})^{n}}{g(n)} \langle I_n \rangle. 	
	\end{align}
	
	The observables with arbitrary $\bar{\beta}$ can be obtained with the relation, 
	\begin{align}
	\langle O(\bar{\beta})\rangle = \frac{\langle \bar{O}(\bar{\beta}) \rangle}{	\langle \bar{Z}(\bar{\beta}) \rangle}.
	\end{align}		
	
	To determine $A$,  we use the fact that the $n = 0$ sector corresponds to a system at infinite temperature ($\beta \rightarrow 0$),
\begin{align}
	A \frac{\langle I_0 \rangle}{g(0)} = 2^{N}, 
\end{align}     
	where $N$ is the total number of  spins in the system. 
Using this relation, the physical partition function, free energy and entropy can be calculated,
	\begin{align}
	\langle Z(\bar{\beta}) \rangle &= \sum_{n} (\bar{\beta})^{n} \frac{I_n}{I_0} \frac{g(0)}{g(n)} 2^{N}, \\
	\langle F(\bar{\beta}) \rangle &= -\frac{1}{\bar{\beta}} \ln \left[ \langle Z(\bar{\beta})\rangle \right], \\
	\langle S(\bar{\beta}) \rangle &= \bar{\beta} \left[ E(\bar{\beta}) - F(\bar{\beta}) \right].
	\end{align}

\section{Topological entanglement entropy and Levin-Wen construction}
 
 As shown in the main text, in order to identify the $Z_2$ QSL, we have to numerically compute the topological entanglement entropy (TEE).
In our simulation, we use the second  Renyi entropy $S_2$ as our entanglement measurement. 
The Renyi entropy with sub-region $A$ follows the area law, 
	\begin{equation}
		S_A= \kappa l - \eta \gamma + O(L^{-1}) 
	\end{equation}   
	where $l$ is the boundary of the sub-region $A$ and $\gamma$ is the topological entanglement entropy. 
	Here, we also consider a finite size correction $O(L^{-1})$ that goes to zero in the thermodynamic limit. 
	
\begin{figure}[ptb]
		\includegraphics[width=\linewidth]{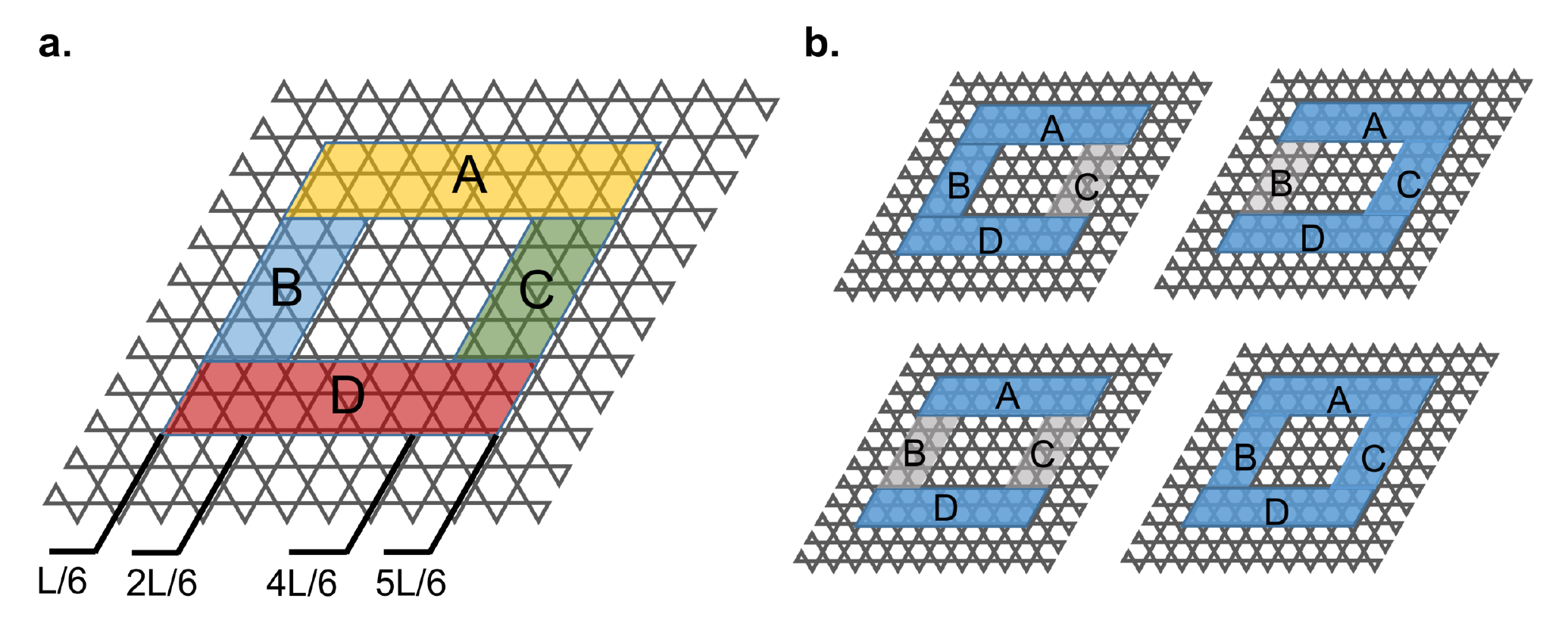}
		\caption{ \textbf{Levin-Wen construction}.  \textbf{a.} Four small parts $A$, $B$, $C$ and $D$ in the system with size $L$. \textbf{b.} Four different sub-regions $R_1$ (upper-left), $R_2$ (upper-right), $R_3$ (lower-left) and $R_4$ (lower-right) are constructed from the four parts $A$, $B$, $C$ and $D$ in order to eliminate the contribution from the boundaries.  }
		\label{FigSuppl:LWSetup} 
	\end{figure}

	The Renyi entropy is computed using QMC  following the procedure in Ref.~\cite{Melko:2010Renyi}.  Application of this method to identify the $Z_2$ topological order can be found in Ref.~\cite{Isakov:2011ee}.
	To estimate the topological entanglement entropy, we use the Levin-Wen construction ~\cite{Wen:2006topo} to eliminate the contributions from the boundaries (area law term). 
	We first construct four different parts out of the lattice; marked by $A$, $B$, $C$ and $D$ as shown in Fig.~\ref{FigSuppl:LWSetup}.

	 We then strategically construct four different sub-regions $R_1$, $R_2$, $R_3$ and $R_4$ with different combination of these four parts as
	\begin{align*}
	R_1 &= A \cup B \cup D,\\
	R_2 &= A \cup C \cup D, \\
	R_3 &= A \cup D, \\
	R_4 &= A \cup B \cup C \cup D.
	\end{align*}
The choice for these subregions allows one to extract the topological entanglement entropy $\gamma$ using the relation
\begin{align*}
2\gamma = -S_2(R_1) - S_2(R_2) + S_2(R_3) + S_2(R_4)
\end{align*}
to eliminate the contributions coming from the boundaries~\cite{Wen:2006topo}.

\section{Hexagon fraction for  $J_{\pm}\ne 0 $ and $J_{\pm\pm}\ne 0$}
\begin{figure}[h!]
	
		\includegraphics[width=0.9\linewidth]{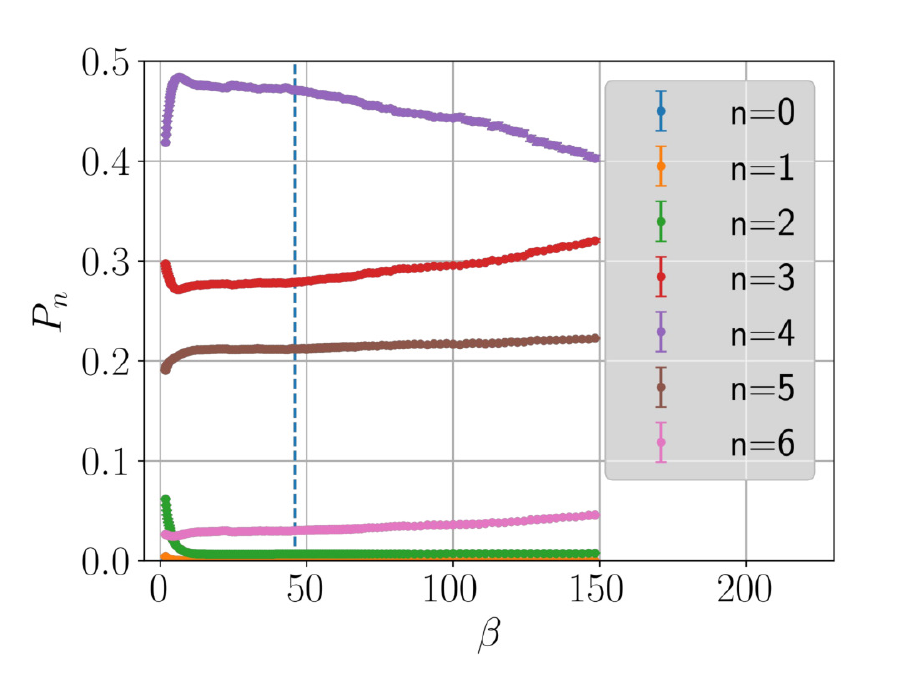}
		\caption{ \textbf{Hexagon occupation fraction}. The hexagon fraction in the VBS region with $h=J_z$, $J_{\pm} = 0.1219J_z$ and $J_{\pm\pm} = -0.25J_z$. The vertical blue line indicates the  perturbative energy scale  }
		\label{FigSuppl:hexamix} 
	\end{figure}

Here we present the hexagon fraction for the case $J_{\pm}\ne 0 $ and $J_{\pm\pm}\ne
0$ where a VBS ground state is expected to establish based on our degenerate
perturbation theory analysis. 	In Fig.~\ref{FigSuppl:hexamix} we show the QMC results
of hexagon fractions $P_n$ at $h=J_z$, $J_{\pm} = 0.1219J_z$ and $J_{\pm\pm} =
-0.25J_z$. The parameters lie in the VBS region with a dominant ring-exchange  term as
the third-order perturbation (as also presented in Fig.~\ref{FigSuppl:Phase3D}b). Our result clearly shows the rise of $P_3$ and the
decrease of $P_4$ at temperature lower than the perturbative energy scale estimated by
the  ring-exchange process $\beta \sim \frac{J_z^2}{12 J_{\pm}^3} = 46/J_z$, with the
same behaviour as in the XXZ model ($J_\pm \ne 0$, $J_{\pm\pm}=0$). This should be contrasted with the behaviour of Fig. 4b in the main text. 
	
\section{Phase diagrams and Structure factors}

\begin{figure*}[ptb]
	\begin{minipage}{0.9\linewidth}
		\includegraphics[width=\linewidth]{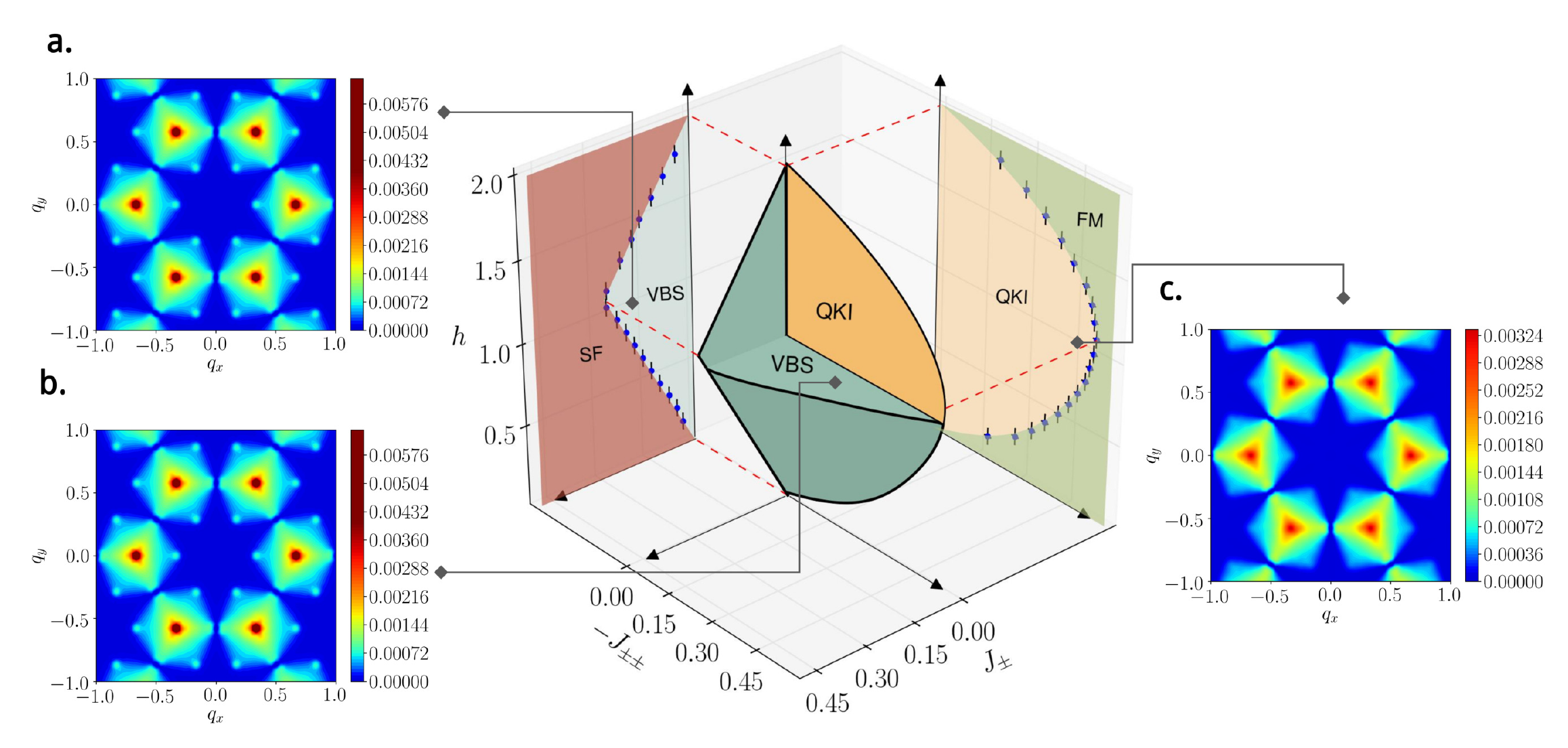}
	\end{minipage}
	\caption{ \textbf{Phase diagram and structure factors}. The schematic phase diagram shows possible phases  of  the XYZh model. 
		The phase boundaries are guides to the eye.
		For $J_{\pm} = 0$ plane, two phases of QKI and ferromagnetic (FM) ordered phase are identified as also shown in Fig.~\ref{FigSuppl:Phase2D}b. For $J_{\pm\pm} = 0$ plane, the lobe of VBS phase appears at $J_{\pm} \ll 1$. The system enters the super fluid phase (SF) when increasing the hopping term $J_{\pm}$ as also shown in Fig.~\ref{FigSuppl:Phase2D}a.
		Structure factors for three cases  with \textbf{a.}  $J_{\pm\pm} = 0$ , $J_{\pm} = 0.19$ and $h = 1.0$,
		\textbf{b.}  $J_{\pm\pm} = -0.25$ , $J_{\pm} = 0.1219$ , $h = 1.0$, and
		\textbf{c.}  $J_{\pm\pm} = -0.45$ , $J_{\pm} = 0$ and $h = 1.0$. 
		In cases \textbf{a} and \textbf{b}, peaks are observed at $\boldsymbol{Q} = \left\langle \frac{2\pi}{3}, 0\right\rangle$ and symmetry related momenta, indicating the emergence of the VBS order; while in the QKI phase (case \textbf{c}), no such peak is observed.
		The structure factors are measured with system size $L=12$ at $T = 0.02$. 
		The momentum vectors $q_x$ and $q_y$ are in unit of $2\pi$ 
		with ferromagnetic peaks being removed for clarity.  
	}
	\label{FigSuppl:Phase3D} 
\end{figure*}

	In Fig.~\ref{FigSuppl:Phase3D} we show a general phase diagram of XYZh model in parameter space $J_{\pm}-J_{\pm\pm}-h$. In the figure, projections to the $J_{\pm}=0$ and $J_{\pm\pm}=0$ planes are shown with the simulation data. 
	To map out the phase boundaries, we take the advantage of the sudden change of the magnetization $M_z$ and magnetic susceptibility $\chi_z$ across the transition  to identify the phase boundaries. The magnetization $M_z$ and magnetic susceptibility $\chi_z$ are defined as
	\begin{align}
		M_z &= \frac{1}{N} \left\langle \sum_{i} S^{z}_i \right\rangle, \nonumber\\
		\chi_z &= \left\langle \left(\sum_{i} S^{z}_i\right)^2 \right\rangle - \left\langle \sum_{i} S^{z}_i\right\rangle^2.
	\end{align}
	
\begin{figure*}[ptb]
\includegraphics[width=\textwidth]{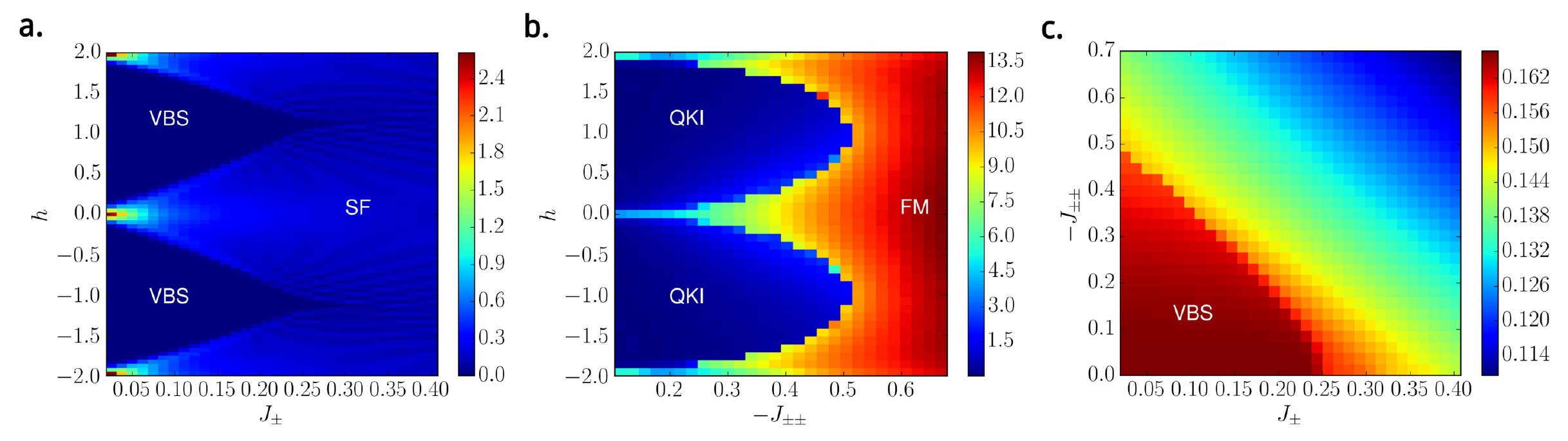}
\caption{ \textbf{Phase diagram in various planes of the parameter space}.  \textbf{a.} Phase diagram in the $J_{\pm}$-$ h$ plane. \textbf{b.} Phase diagram in the $J_{\pm\pm} - h$ plane. \textbf{c.} Phase diagram in the $J_{\pm} - J_{\pm\pm}$ plane with $h = 1$. Simulations are performed  with system size $L=6$ at temperature $T = 0.015J_z$ using standard SSE. For \textbf{a.} and \textbf{.b}, the phase diagram are mapped by the magnetic susceptibility $\chi_z$. For \textbf{(c)}, the phase diagram is mapped by the magnetization $M_z$.}
\label{FigSuppl:Phase2D} 
\end{figure*}

	Fig.~\ref{FigSuppl:Phase2D} shows the phase diagrams of various cross section of the parameter space. For plane with $J_{\pm}=0$, two phases of QKI and ferromagnetic (FM) are identified, which is consistent with previous study ~\cite{Hao:2015qki}. For plane with $J_{\pm\pm}=0$, we have VBS and superfluid (SF) phase as ~\cite{Isakov:2006vbs}.
	
	At $J_{\pm}-J_{\pm\pm}$ plane with a horizontal cross section at $h = 1.0$, we find a lobe with VBS ordering at a finite $J_{\pm}$. The emergence of VBS is consistent and expected as a consequences of introducing a third order ring-exchange term that is shown in our DPT analysis.    
		
\begin{figure}[ht!]
	\includegraphics[width=\linewidth]{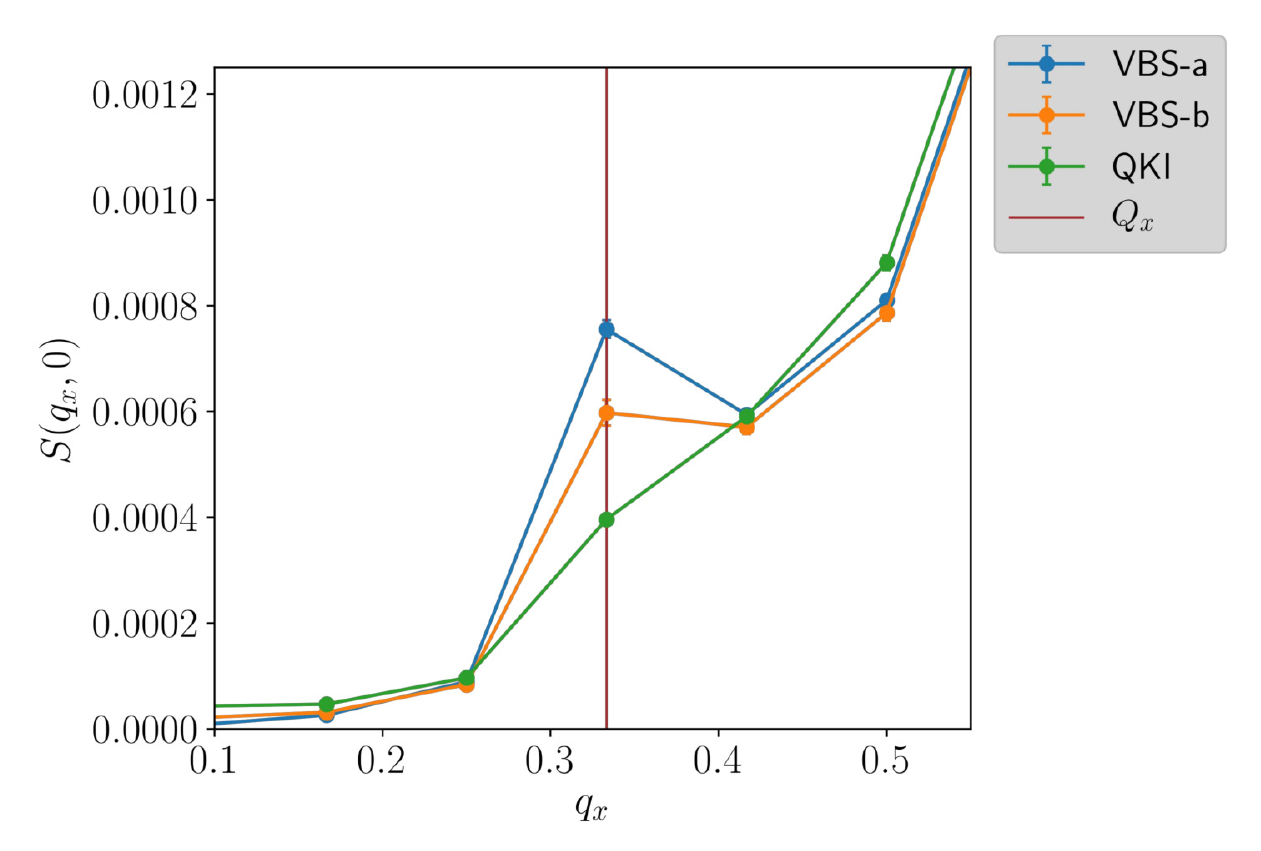}
	\caption{ \textbf{Structure factors in various phases}. The structure factors are measured at $N=3\times 12\times 12$ at $T = 0.02$. 
		The curves of VBS-a, VBS-b and QKI correspond to the line cut along $\mathbf{q}=(q_x, 0)$ of Fig.~\ref{FigSuppl:Phase3D}a, Fig.~\ref{FigSuppl:Phase3D}b and Fig.~\ref{FigSuppl:Phase3D}c respectively. The $x$ component momentum vector $q_x$ are in units of $2\pi$. $Q_x=\frac{2\pi}{3}$ indicates the VBS ordering vector.	
		}
		\label{FigSuppl:SFs} 
	\end{figure}

	The three-fold degenerate VBS state with broken translational symmetry can be identified from the peaks of the static structure factor $S(\mathbf{q})$ at ordering momentum vector $\mathbf{q}=\left\langle\frac{2\pi}{3},0\right\rangle$ and symmetry related momenta~\cite{Isakov:2006vbs}. The static structure factor defines as :

	\begin{align}
	f(\mathbf{q}) &= \frac{1}{N} \sum_{j} e^{i\mathbf{q} \cdot \mathbf{r}_j} S^{z}_j \nonumber\\
	S(\mathbf{q})   &= \left\langle f(\mathbf{q})f(-\mathbf{q})\right\rangle - \left\langle f(\mathbf{q})\right\rangle \left\langle f(-\mathbf{q})\right\rangle
	\end{align}
	with $N = 3 \times L \times L$ is the total number of spins. Fig.~\ref{FigSuppl:SFs} shows the line cut along $\mathbf{q}=(q_x, 0)$ of the structure factors shown in Fig.~\ref{FigSuppl:Phase3D}.  In both the VBS-a and VBS-b cases, peaks at $\mathbf{q}=\left\langle\frac{2\pi}{3},0\right\rangle$ emerge out of the background.

\section{Ground states of the modified effective model}

	To understand the ground states of the modified effective model, we analyse the spectral properties of the energy eigenstates $\ket{\phi_i}$ obtained by exact diagonalisation. 
	We write the state of interest  $\ket{\phi_i}$ in terms of the classical kagome ice basis $\{ \Omega_n \}$  where the energy eigenstate $\ket{\phi_i}$ can be represented as:
	
\begin{equation}
	\ket{\phi_i} = \sum_{n} A_n \ket{\Omega_n}, \quad C_n \equiv \left|A_n\right|^2,
\end{equation}
	where $C_n$ corresponds to the probability of the classical state $\Omega_n$.

	We first study the effective model in the  classical limit with only the diagonal term Eq.~(\ref{diag}) present. 	This amounts to taking $\alpha\to\infty$ in the original $\hat{P}_6$ model in the main text.
 
\begin{figure}[ht!]
	\includegraphics[width=\linewidth]{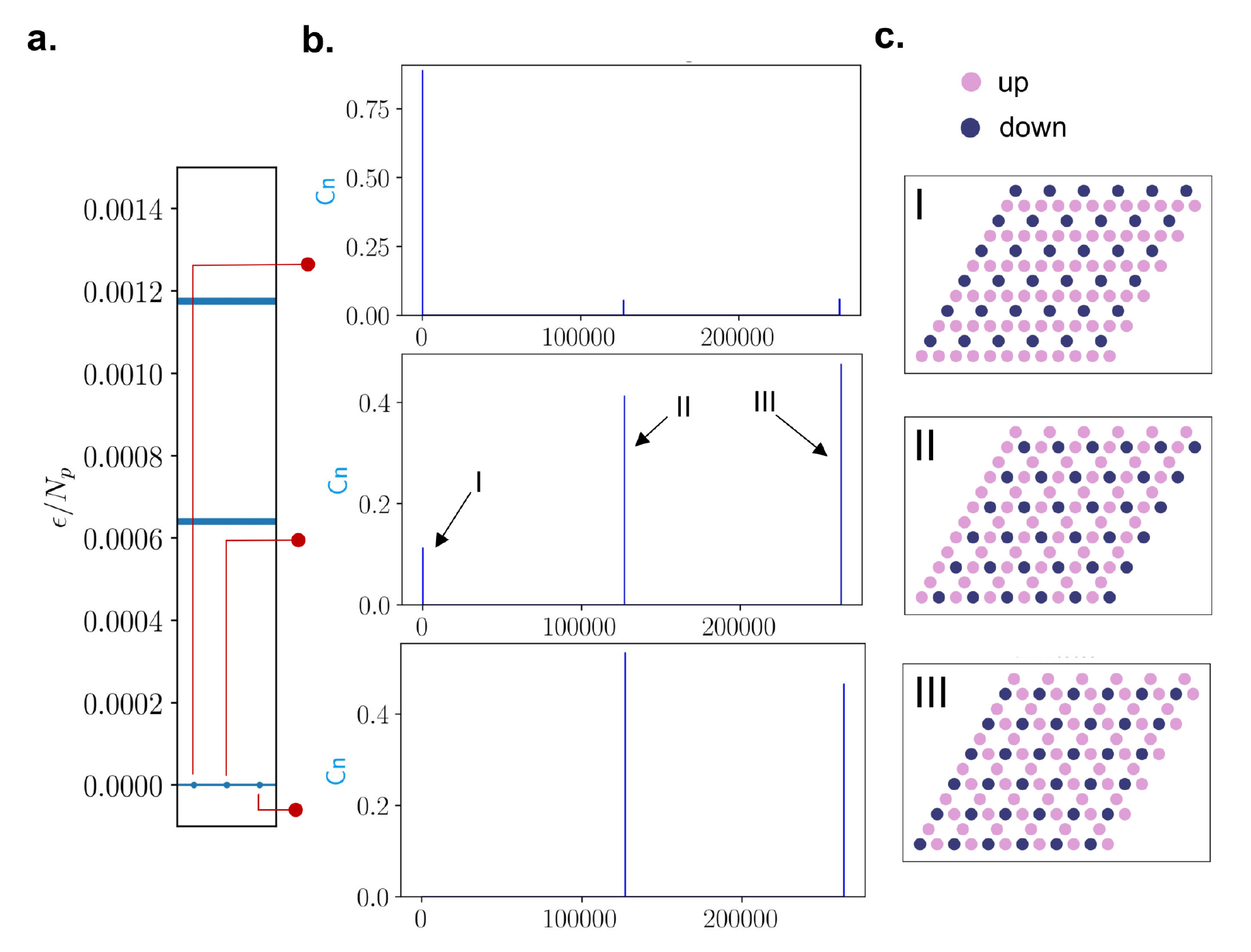}
\caption{ \textbf{Energy spectrum of  the diagonal-only model}. \textbf{a.} The   spectrum  of the effective model (\ref{diag}) with only the diagonal terms  for $N = 6 \times 6 \times 3$,  $J_{\pm\pm}=-0.49, h=1.0, J_{\pm}=0$. $N_p=N/3$ is the number of hexagons. The ground energy is shifted to zero. \textbf{b.} The three-fold degenerate ground states represented in the classical kagome ice states.  The blue bar represents the probability $C_n$ for each configuration and the $x$-axis is the classical configuration index.where $x$-axis is the classical configuration index. The ground states corresponds to linear combination of three charge-ordered states in classical kagome ice \textbf{c.}.  }	
\label{FigSuppl:L6Donly}
\end{figure}

Fig.~\ref{FigSuppl:L6Donly}a shows the energy spectrum and we find 
the ground states are three-fold degenerate.
These states are linear combination of three possible
charge-ordered states in the classical kagome ice~\cite{Chern:2011fv,Wolf:CIKag}
(Fig.~\ref{FigSuppl:L6Donly}b), marked with $\mathrm{I}$, $\mathrm{II}$ and $\mathrm{III}$ shown in Fig.~\ref{FigSuppl:L6Donly}c. 
Note that every hexagons within these 
charge-ordered 
 configurations are all $n=4,b$. 
These states are smoothly connected to the ground states for $\alpha\gtrsim 1.703$.

\begin{figure}[ht!]	
\includegraphics[width=\linewidth]{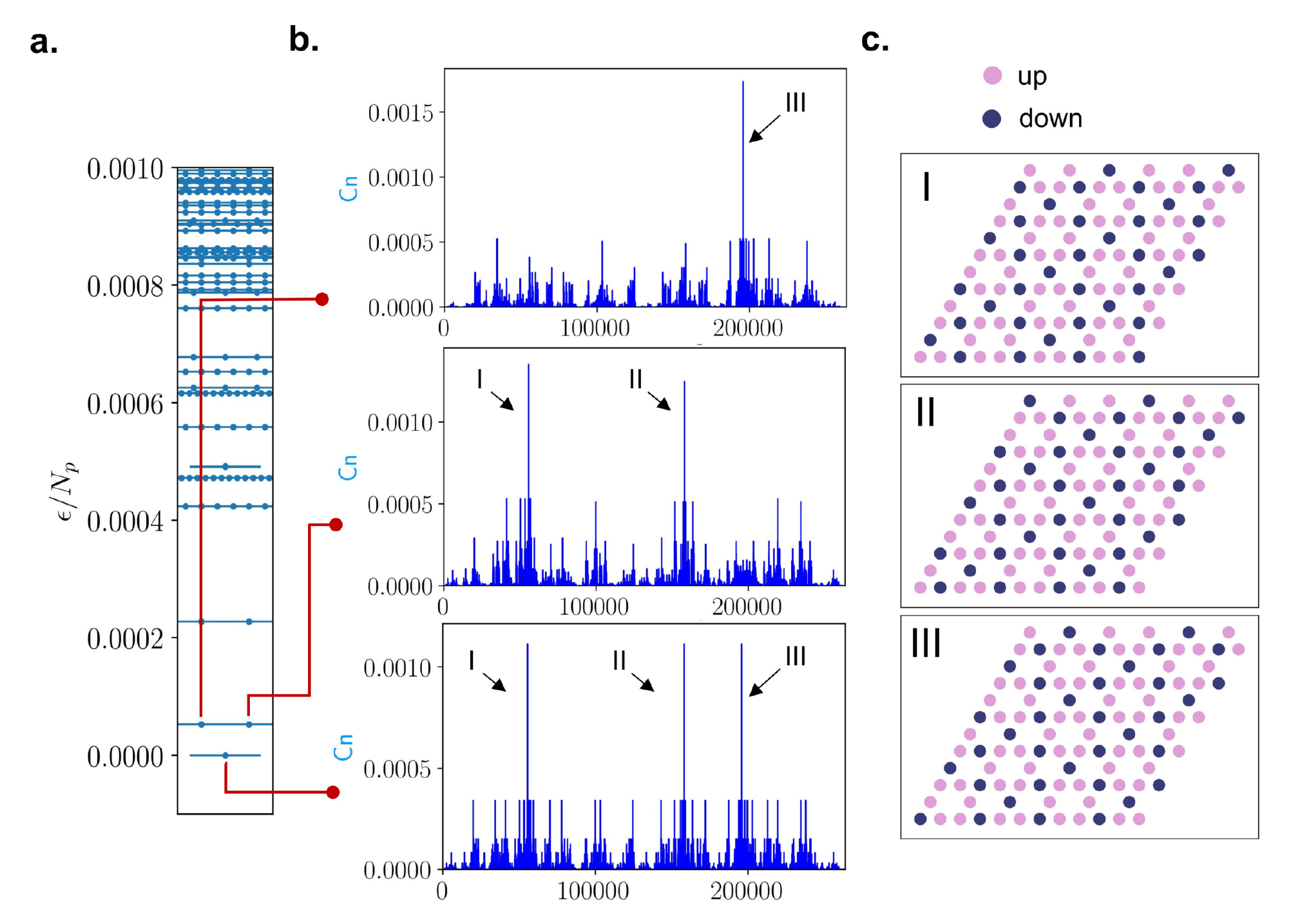}
\caption{\textbf{Energy spectrum of the ring-exchange  model}. \textbf{a.} The spectrum  of the ring-exchange model Eq.~(\ref{offdiag}). The ground energy is shifted to zero. The system size is $N= 6\times 6\times 3$ and $N_p=N/3$ is the number of hexagons. \textbf{b.} The three lowest energy eigenstates represented in the classical kagome ice states. The blue bar represents the probability $C_n$ for each configuration and the $x$-axis is the classical configuration index. \textbf{c.} Three  ice configurations with  dominant probabilities $C_n$ for the three lowest energy states. The classical configurations correspond to the  $\sqrt{3} \times \sqrt{3}$ states. The parameters considered here are $J_{\pm\pm}=-0.49, h=1.0, J_{\pm}=0$.   
		}
\label{FigSuppl:L6ODonly}
\end{figure} 	
	
\begin{figure}[hb!]
	\includegraphics[width=0.65\linewidth]{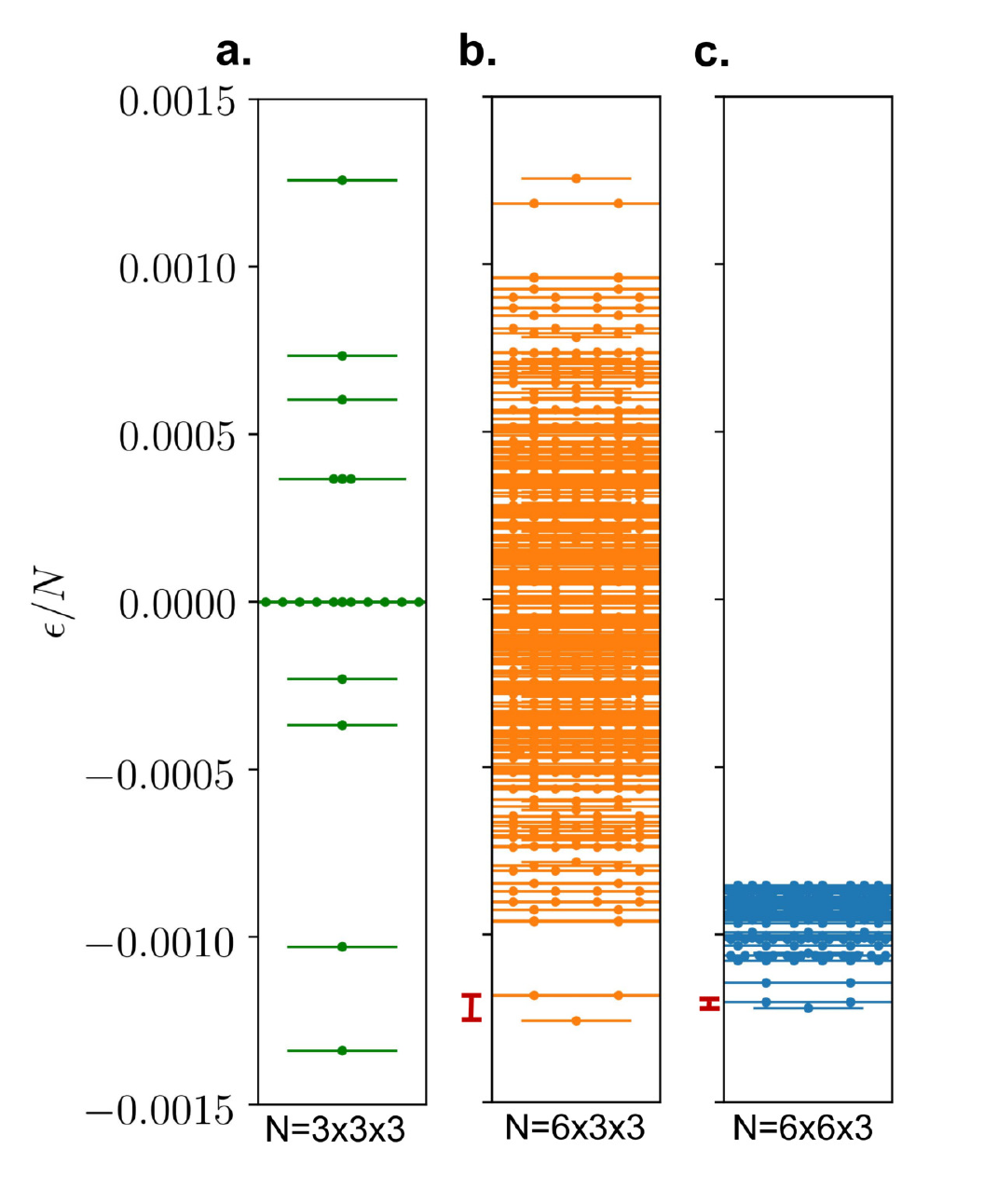}	
	\caption{ \textbf{The energy spectrums of different system sizes}. The full  spectrum the effective model at $\alpha=0$ for systems with total number of spins $N = 3 \times 3 \times 3$ (\textbf{a})  , and $N = 6 \times 3 \times 3$ (\textbf{b}) , and the lowest 300 states for system  $N = 6 \times 6 \times 3$ (\textbf{c}).  The finite-size gap between the ground states and the first-excited state, indicated by the red line,  decreases as the system size increases. The parameters are $J_{\pm\pm}=-0.49, h=1.0, J_{\pm}=0$.}
	\label{FigSuppl:FSODonly}
\end{figure}

In the other  limit $\alpha = 0$, where only the ring-exchange term Eq.~(\ref{offdiag}) is present, we expect a three-fold degenerate VBS ground state~\cite{Isakov:2006vbs,Moessner:QDM} in the thermodynamic limit. 
In a finite-size simulation, these states are not  exactly degenerate. 
However, the spectral property of these states should manifest the VBS signature.  
Fig.~\ref{FigSuppl:L6ODonly}a shows the energy spectrum of the ring-exchange model.   
The three lowest energy states 
( Fig.~\ref{FigSuppl:L6ODonly}b)  are the linear superposition of classical configurations with $n=3$ hexagons, dominated by three configurations corresponding configurations to the $\sqrt{3}\times\sqrt{3}$ states as shown in Fig.~\ref{FigSuppl:L6ODonly}c.
The VBS states  are generated by  the tunnelling between these three $\sqrt{3}\times\sqrt{3}$ states with the ring-exchange, and other $n=3$ states are intermediate configurations generated from the tunnelling processes.
These states are smoothly connected to the lowest energy states for $\alpha\lesssim1.703$.

Finally, to show the three lowest states will eventually combined to form the three-fold degenerate states in the thermodynamic limit, we compare the finite-size gap for different system sizes as shown in Fig.~\ref{FigSuppl:FSODonly}.
 We see that the gap between the lowest three states decreases as system size increases, and the states become degenerate in the thermodynamic limit.

\clearpage

\bibliography{refQKIv2}

\end{document}